\documentclass[final,3p,twocolumn,12pt,authoryear,fleqn]{elsarticle}

\usepackage{amssymb,amsthm,booktabs}
\usepackage{astro}
\usepackage[version=3]{mhchem}

\journal{Icarus}

\begin{document}

\setlength{\mathindent}{0pt}
\def\perth{\ensuremath{\hbox{$\,^0\!/_{00}$}}}
\def\iras{IRAS16293-2422}
\def\mc#1#2{\multicolumn{#1}{c}{#2}}
\def\Ntot   {\ensuremath{N_{\rm tot}}}
\def\atoratio  {\ce{^{14}N}/\ce{^{15}N}}
\def\ratio#1{\ensuremath{{\cal R}_{\ce{#1}}}}  
\def\fourn  {\ensuremath{{\rm ^{14}N}}}
\def\fifn   {\ensuremath{{\rm ^{15}N}}}
\def\foun{\ensuremath{\ce{$^{14}$N}}}
\def\nhhd   {\ensuremath{\rm NH_2D}}
\def\nhhh   {\ensuremath{\rm NH_3}}
\def\fifnh  {\ensuremath{{\rm ^{15}NH}}}
\def\fifnhh {\ensuremath{\rm ^{15}NH_2}}
\def\fifnhhh{\ensuremath{\rm ^{15}NH_3}}
\def\fifnnhp{\ensuremath{\rm ^{15}NNH^+}}
\def\nfifnhp{\ensuremath{\rm N^{15}NH^+}}
\def\NH     {\ensuremath{\rm NH}}
\def\NHH    {\ensuremath{\rm NH_2}}
\def\founh  {\ensuremath{\rm ^{14}NH}}
\def\comm#1 {\textbf{\Ra #1}}
\def\dearth{\ensuremath{\delta_{\oplus}}}
\def\dearth{\ensuremath{\delta{\fifn}}}
\def\dsun{\ensuremath{\delta_{\odot}}}
\def\corr#1{{#1}}

\begin{frontmatter}



  \title{The \fifn-enrichment in dark clouds and Solar System objects}

\author{Hily-Blant P., Bonal L., Faure A., Quirico E.}
\address{Universit\'e Joseph Fourier/CNRS, Institut de
  Plan\'etologie et d'Astrophysique de Grenoble, France}


\begin{abstract}
  
  The line intensities of the fundamental rotational transitions of
  \ce{H$^{13}$CN} and \ce{HC$^{15}$N} were \corr{observed} towards two
  prestellar cores, L183 and L1544, and lead to molecular isotopic
  ratios $140\le \atoratio \le 250 $ and $140 \le \atoratio \le 360$,
  respectively. The range of values reflect genuine spatial variations
  within the cores.  A comprehensive analysis of the available
  measurements of the nitrogen isotopic ratio in prestellar cores show
  that molecules carrying the nitrile functional group appear to be
  systematically \fifn-enriched compared to those carrying the amine
  functional group.  A chemical origin for the differential
  \fifn-enhancement between nitrile- and amine-bearing interstellar
  molecules is proposed. This sheds new light on several observations
  of Solar System objects: \textit{(i)} the similar N isotopic
  fractionation in Jupiter's \ce{NH3} and solar wind \ce{N^+};
  \textit{(ii)} the \fifn-enrichments in cometary HCN and CN (that
  might represent a direct interstellar inheritance); and
  \textit{(iii)} \fifn-enrichments observed in organics in primitive
  cosmomaterials.  The large variations in the isotopic composition of
  N-bearing molecules in Solar System objects might then simply
  reflect the different interstellar N reservoirs from which they are
  originating.

\end{abstract}

\begin{keyword}
  astrochemistry \sep cosmochemistry \sep Solar Nebula \sep meteorites
  \sep Origin Solar System \sep radio observations \sep prestellar cores
  \sep objects: L1544, L183


\end{keyword}

\end{frontmatter}

\section{Introduction}

Nitrogen, the fifth most abundant element in the Universe, exists
naturally as a highly volatile gas (\ce{N2}, N) and a mixture of
compounds of varying volatility (such as \ce{NH3}, HCN, HNC,
\etc). The relative abundances and isotopic compositions of these
different nitrogen occurrences in various astronomical sources can
provide useful clues to the origin and history of the Solar System.

The Sun formed from a cold and dense core embedded in its parental
interstellar molecular cloud rich in gas and dust. The so-called
``protosolar nebula'' (PSN) is the evolutionary stage issued from the
collapsing prestellar core. The nitrogen volatile isotopologues in
this nebula may have been fractionated with respect to the original
interstellar material, \ie\ the isotopic ratio measured in these
molecules may differ from the elemental ratio. Such fractionation
processes are invoked to explain the large enhancements of the D/H
ratio measured in several molecular species in prestellar cores
\citep[\eg\ ][]{caselli2003, roueff2005}. The efficiency of these
processes however depends on the physical conditions in the core
during its collapse \citep{flower2006a}. One of the current challenges
in astrochemistry is to follow the chemical composition of a starless
core during its evolution towards a planetary system. The related
challenge in cosmochemistry is to identify, in primitive objects of
the Solar System, residual materials from the original cloud.

The Sun is the largest reservoir of nitrogen in the Solar
System. Isotopic measurements of solar wind trapped in lunar soils
\citep{hashizume2000}, analysis of Jupiter's atmosphere
\citep{fouchet2000, owen2001} and osbornite (TiN), considered as the
first solid N-bearing phase to form in the cooling protosolar nebula
\citep{meibom2007}, all independently showed that nitrogen in the PSN
was much poorer in \fifn\ than the terrestrial atmosphere. The
analysis of the present-day solar wind trapped on Genesis targets
finally concluded on and confirmed these previous studies. The solar
wind is depleted in \fifn\ relative to inner planets and meteorites,
and define the following atomic composition for the present-day Sun
$\atoratio = 441\pm5$ \citep{marty2010, marty2011}. The isotopic
composition of nitrogen in the outer convective zone of the Sun has
not changed through time and is considered as representative of the
PSN.  In the present paper, we only consider the original/primary N
isotopic fractionation, as opposed to secondary \fifn-enrichments
acquired through atmospheric process (\eg\ Titan, Mars) for
example. In the remainder of the paper and for the sake of clarity,
the elemental isotopic ratio is noted \atoratio, whilst the isotopic
ratio X\fifn/X\foun\ measured in any N-bearing species X is noted
\ratio{X}.

\begin{figure*}
  \centering
  \includegraphics[width=\hsize]{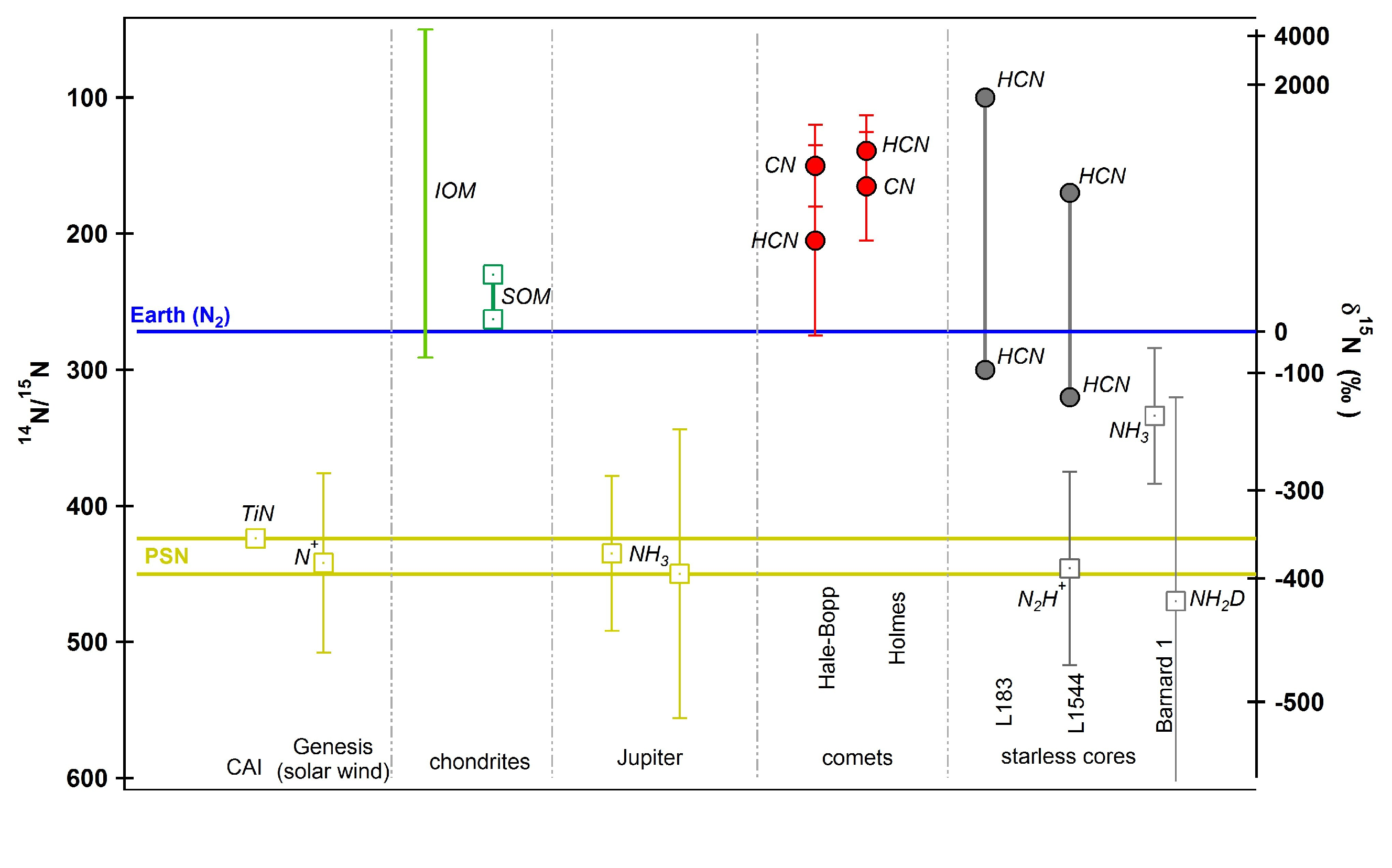}
  \caption{Nitrogen isotopic composition of Solar System objects as
    compared to the composition of simple molecules in interstellar
    clouds. The isotopic composition is expressed in term of
    \atoratio\ ratios (left scale) and in $\dearth$ notation (right
    scale, $\dearth = [\ratio{\oplus}/\ratio{} -1 ]\times 1000$, where
    $\ratio{\oplus} = 272$ is the nitrogen isotopic composition of the
    terrestrial atmosphere, see also Table~\ref{tab:all}). Square and
    circle symbols are for measurements made on molecules with amine
    and nitrile functional groups, respectively. IOM stands for
    Insoluble Organic Matter, SOM for Soluble Organic Matter, and CAI
    for Calcium-, Aluminum-rich Inclusions. The range of values
    reported towards L183 and L1544 reflect the spatial variations
    accross the sources.}
  \label{fig:ssratio}
\end{figure*}

In our Solar System, any object (with the exception of Jupiter) is
actually enriched in \fifn\ compared to the PSN (see
Fig.~\ref{fig:ssratio}).  Large excesses in \fifn\ have been found in
organic material of chondrites and interplanetary dust particles
(IDPs). Enrichments in \fifn\ are measured at different scales of the
material (bulk \vs\ hotspots) and can be as high as $\ratio{} = 50$
\citep{messenger2000, bonal2010}. Molecules in cometary coma also
appear to be \fifn-enriched, with \ratio{}\ ratios varying between 139
and 205 in HCN and CN \citep[see the review by][]{jehin2009}.

The variation of the nitrogen isotopic composition in Solar System
objects is \corr{most likely} caused by a variety of effects. These
include : \textit{(i)} nucleosynthetic origin \corr{\citep[][and
  references therein]{audouze1985, adande2012}}; \textit{(ii)}
photochemical self-shielding in the solar nebula
\corr{\citep{clayton2002, lyons2009}}; \textit{(iii)} spallation
reactions caused by the irradiation of the young sun
\corr{\citep{kung1978, chaussidon2006}}; \textit{(iv)} low temperature
isotope exchanges \corr{\citep[][hereafter TH00]{terzieva2000}}. The
absence of large \fifn-enrichments accross the Galaxy \citep[][ and
references therein]{adande2012} and the small fractionation effects
predicted by standard gas-phase chemical models (TH00) have weakened
so far the hypothesis of a preserved (low temperature) interstellar
chemistry to explain the \fifn-enrichments observed in primitive solar
cosmomaterials. \corr{However, the gas-grain chemical model of
  \cite{charnley2002} is able to reach a significant \fifn\ enrichment
  of ammonia which is eventually locked into ices}. The absence of a
direct correlation between D and \fifn-enrichments in organics from
primitive cosmomaterials has also been interpreted as the lack of
remnant interstellar chemistry for N isotopologues
\corr{\citep{briani2009, marty2010, aleon2010}}.

In the present work, we analyze the line intensities of the
fundamental rotational transitions of \ce{H$^{13}$C$^{14}$N} and
\ce{H$^{12}$C$^{15}$N} (\hthcn\ and \hcfifn\ in the following) towards
two starless dense cores, L1544 and L183 \citep{hilyblant2010n}. The
main novelty in our analysis stems from the recent availability of
accurate collisional hyperfine selective rate coefficients for HCN
with \hh\ \citep{benabdallah2012}. The present work \textit{(i)}
brings new observational constraints on nitrogen isotopic
fractionation in gas phase and \textit{(ii)} puts a new perspective on
the actively debated and long questioning issue of the origin of the
\fifn-enrichments observed in primitive cosmomaterials as compared to
the protosolar nebula.

\section{Material and methods}
\label{sec:material}

\subsection{Observations}

Observations of the pure rotational $J=1-0$ lines of \hthcn\ and
\hcfifn\ were carried out with the IRAM-30m telescope by
\cite{hilyblant2010n}. Spectra along perpendicular directions towards
the L183 and L1544 starless cores were obtained, with extremely high
spectral resolution ($\nu_0 / \delta \nu \approx 4\tdix{6}$), such
that the hyperfine structure of the \hthcn(1-0) is resolved. The
details of the observational setup and hardware performances are
available in \cite{hilyblant2010n}. The \hthcn\ and \hcfifn(1-0)
spectra towards L183 and L1544 are shown in
Fig.~\ref{fig:spectral183}. The data are analyzed following a more
robust method than the one previously adopted, where column densities
were derived under the Local Thermal Equilibrium (LTE) assumption at a
temperature of 8~K. In the present analysis, we make use of the
hyperfine structure of the \hthcn(1-0) line and of new collisional
coefficients for HCN-\hh\ \citep{benabdallah2012} which were also
adopted for \hthcn\ and \hcfifn.

\subsection{Data analysis}

\begin{table*}
  \centering
  \caption{Column densities (\tdix{12}\cc) of \ce{H^{13}CN}(1-0) from three methods,
    and of \hcfifn\ (LVG calculation) towards L183.}
  \begin{tabular}{c c c c c c c c c c c c c c c c}
    \toprule
    \mc{2}{(1)} & (2)
    & (3) & (4) & (5)
    & (6) & (7) & (8)
    & (9) & (10)& (11) 
    & (12) & (13)
    \\
    \mc{2}{Offset} & FWHM
    & $\tau_0$ & \texc & $N_{13}$ 
    & $\tau_0$ & \texc & $N_{13}$ 
    & $N_{13}$ 
    & $N_{15}$ & $\frac{N_{13}}{N_{15}}$ & \ratio{HCN} & \dearth
    \\ 
    \mc{2}{arcsec} & \kms & & K & \cc & & K & \cc & \cc & \cc  \\
    \midrule
  -40&  0 & 0.36 &  2.3 & 2.9 & 1.8 &  3.1 & 2.8 & 2.3 &  2.3 & --  & &\\
  -20&  0 & 0.42 &  1.6 & 3.2 & 1.7 &  1.8 & 3.0 & 1.7 &  1.7 & --  & &\\
    0&-40 & 0.34 &  3.7 & 3.0 & 2.9 &  3.4 & 2.9 & 2.5 &  2.7 & --  & &\\
    0&-20 & 0.33 &  3.2 & 3.1 & 2.5 &  3.1 & 2.9 & 2.2 &  2.6 & 0.9 & 3.1 &208 &306\\
    0&  0 & 0.47 &  4.4 & 3.0 & 4.7 &  4.5 & 2.8 & 4.4 &  4.7 & 1.3 & 3.7 &252 & 79\\
    0& 20 & 0.42 &  3.7 & 3.0 & 3.5 &  3.1 & 2.9 & 2.8 &  3.3 & 1.1 & 2.9 &197 &382\\
    0& 40 & 0.36 &  3.5 & 3.1 & 3.0 &  2.6 & 2.9 & 2.0 &  2.3 & 1.1 & 2.0 &136 &995\\
   20&  0 & 0.43 &  1.7 & 3.1 & 1.7 &  0.5 & 3.2 & 0.5 &  0.5 & --  & &\\
   40&  0 & 0.46 &  0.3 & 3.8 & 0.4 &  0.8 & 3.0 & 0.8 &  0.9 & --  & &\\
   \bottomrule
  \end{tabular}
  \label{tab:l183}
  \begin{list}{}
    \scriptsize
  \item\textbf{Notes}:
  \item (1): spatial offsets with respect to $(\alpha,\delta)_{J2000}
    = (15^h 54^m 08.80^s, -02\degr 52' 44.0\arcsec)$.
  \item (2): line width (assumed identical for the three components)
    from 3-components Gaussian fits, for the 3 hyperfine components.
  \item (3), (4), (5): total center line opacity, excitation
    temperature, and total \hthcn\ column density (\dix{12}\,\cc) in
    LTE as deduced from a HFS fit in CLASS (see text).
  \item (6), (7), (8): same as above but as derived from the
    3-components Gaussian fits.
  \item (9): the column density is calculated in the LVG
    approximation, from a $\chi^2$-minimization against the opacity
    $\tau_0$ and the line intensity of the strongest hyperfine
    component. The column densities only weakly vary with the kinetic
    temperature in the range 5 to 10~K. The values here correspond to
    $\tkin=8$~K.
  \item (10): column density of \hcfifn\ calculated under the LVG
    approximation for the density and kinetic temperature
    corresponding to the best solution from the \hthcn\ LVG
    calculations.
  \item (11), (12), (13): column density ratios and isotopic ratios
    assuming HCN/\hthcn=68. $\dearth = [\ratio{\oplus}/\ratio{} -1
    ]\times 1000$, where $\ratio{\oplus} = 272$ is the nitrogen
    isotopic composition of the terrestrial atmosphere
  \end{list}
\end{table*}

\begin{table*}
  \centering
  \caption{Column densities (\tdix{12}\cc) of \ce{H^{13}CN} and \hcfifn\ towards L1544.}
  \begin{tabular}{c c c c c c c c c c c}
    \toprule
    \mc{2}{(1)} & (2)
    & (3) & (4) & (5)
    & (6) & (7) & (8)
    & (9) & (10)
    \\
    \mc{2}{Offset} & FWHM
    & $\tau_i$ & \texc & \nhh & $N_{13}$ & $N_{15}$ & $\frac{N_{13}}{N_{15}}$ & \ratio{HCN} & \dearth\\
    \mc{2}{arcsec} & \kms & & K & \dix{4}\ccc & \cc & \cc\\
    \midrule
    -40 &  40 &  0.53 &  2.50 & 3.1 &  6.3 & 5.3 & 1.7 &  3.2 & 215 &  266\\
    -20 & -20 &  0.40 &  1.28 & 3.3 & 12.5 & 2.5 & 0.6 &  4.5 & 309 & -119\\
    -20 &  20 &  0.59 &  1.86 & 3.2 & 10.0 & 4.7 & 1.1 &  4.4 & 296 &  -81\\
      0 &   0 &  0.46 &  1.83 & 3.5 & 12.5 & 4.6 & 1.2 &  3.8 & 257 &   58\\
     20 & -20 &  0.41 &  2.16 & 3.3 & 10.0 & 4.1 & 1.2 &  3.5 & 238 &  142\\
     20 &  20 &  0.42 &  1.52 & 3.1 & 10.0 & 2.6 & 0.9 &  3.0 & 207 &  316\\
     40 & -40 &  0.30 &  1.68 & 3.2 & 10.0 & 2.4 & 1.2 &  2.0 & 136 &  993\\
    \bottomrule
  \end{tabular}
  \label{tab:l1544}
  \begin{list}{}
    \scriptsize
  \item\textbf{Notes}:
  \item (1): spatial offsets with respect to $(\alpha,\delta)_{J2000}
    = 05^h04^m16.90^s, 25\degr10'47\arcsec)$.
  \item (2): line width (assumed identical for the three hyperfine
    components) from independent Gaussian fits.
  \item (3), (4): \hthcn(1-0) center line opacity of the hyperfine
    component with relative intensity RI=0.5556, and excitation
    temperature, derived from the relative integrated intensities of
    the three hyperfine components, assuming equal \texc\ for the
    three hyperfine components.
  \item (5), (6), (7): \hh\ density and total \hthcn\ and \hcfifn\
    column densities, derived through $\chi^2$-minimization accross
    LVG calculations. Minimization is done in the \nhh, \tkin\ plane
    using the \hthcn\ opacity and line intensity of the RI=0.5556
    component as constraints. The column density of \hcfifn\ derives
    from LVG calculations at the \nhh, \tkin\ given by \hthcn. The
    values here correspond to $\tkin=8$~K.
  \item (8), (9), (10): column density ratios and isotopic ratios
    assuming HCN/\hthcn=68.
  \end{list}
\end{table*}

The analysis of the data makes use of the hyperfine structure of
\hthcn. The total opacity and excitation temperature of the
\hthcn(1-0) transition are derived, assuming equal excitation
temperature within the hyperfine multiplet. This assumption is
justified as long as the opacity remains of the order of unity, which
as will be seen later, holds for the lines towards L1544 and L183. The
opacity and excitation temperature may then be used to derive the
column densities under the LTE assumption (see details in the
Appendix). Alternatively, the opacity and line intensity may serve to
compute the column density, \hh\ number density, and kinetic
temperature, from non-LTE calculations, under the so-called Large
Velocity Gradient framework. In such case, we have used the RADEX
public code \citep{vandertak2007}. In these calculations, the \hthcn\
column density is searched for by varying the \hh\ density and the
kinetic temperature in the range \dix{11} to \dix{14}\cc, \dix{3} to
\dix{7}\ccc, and 5 to 15~K, respectively.


In the case of L183, three methods have been compared. \textit{1/} The
HFS method from the CLASS software was applied (see Appendix) with the
opacity and the excitation temperature as outputs, which in turn serve
to compute a LTE column density. \textit{2/} Another fitting method
was based on three independent Gaussians, yet constrained to have the
same linewidth, whose peak intensities were used to derive the opacity
and the excitation temperature. These two outputs give another LTE
estimate of the total column density. \textit{3)} The opacity and line
intensity of a given hyperfine component (\eg\ the one with RI=0.5556)
from the latter fitting method were used to derive the column density
from LVG calculations. The results of these three methods are
summarized in Table~\ref{tab:l183}.

The case of L1544 was tackled in a slightly different fashion, to
handle the double peak line profiles, which likely result from two
different velocity components along the line of sight rather than
infall, as these double peaks are seen in both optically thin and
thick tracers. Each hyperfine component was thus fitted as two
independent Gaussian profiles, from which an integrated intensity and
equivalent linewdith are derived. The relative integrated intensites
are used to estimate the opacity and excitation temperature (see
Eq.~\ref{eq:hfs1}). Finally, the opacity and integrated intensity are
$\chi^2$-minimized in the \{\nhh, $N$(\hthcn)\} plane through LVG
calculations, at various kinetic temperatures.

The \hcfifn\ column density was obtained from LVG calculations using
the \hcfifn(1-0) line intensity as a constraint.  Solutions in terms
of the \hcfifn\ column density are thus obtained by matching the LVG
predictions to the observed intensities. Because the hyperfine
structure of \hcfifn\ is not resolved out, we adopted the physical
conditions derived from the LVG \hthcn\ analysis while varying only
the \hcfifn\ column density. This assumes that the two molecules
coexist spatially, which is a reasonable assumption based on simple
chemical considerations which show that both molecules derive from the
same chemical paths \citep[\eg\ TH00, ][]{hilyblant2010n}. The
signal-to-noise ratio of the \hcfifn\ spectra towards L183 was found
to be good enough for only 4 positions. The results of these
calculations are given in Tables~\ref{tab:l183} and
\ref{tab:l1544}. The corresponding isotopic ratios are shown in
Fig.~\ref{fig:observed-ratio}. The typical statistical uncertainty on
the derived column densities is 10\%. Towards L183, the comparison of
the column density resulting from the three methods provide a more
reliable estimate of the uncertainty on the column density
determination, of the order of 20\%. Towards L1544, we have used
separately the RI=0.3333 and RI=0.5556 line intensities as
constraints, which results in a dispersion of 10 to 30\%.

\subsection{Results}

\begin{figure*}
  \centering
  \includegraphics[width=\hsize]{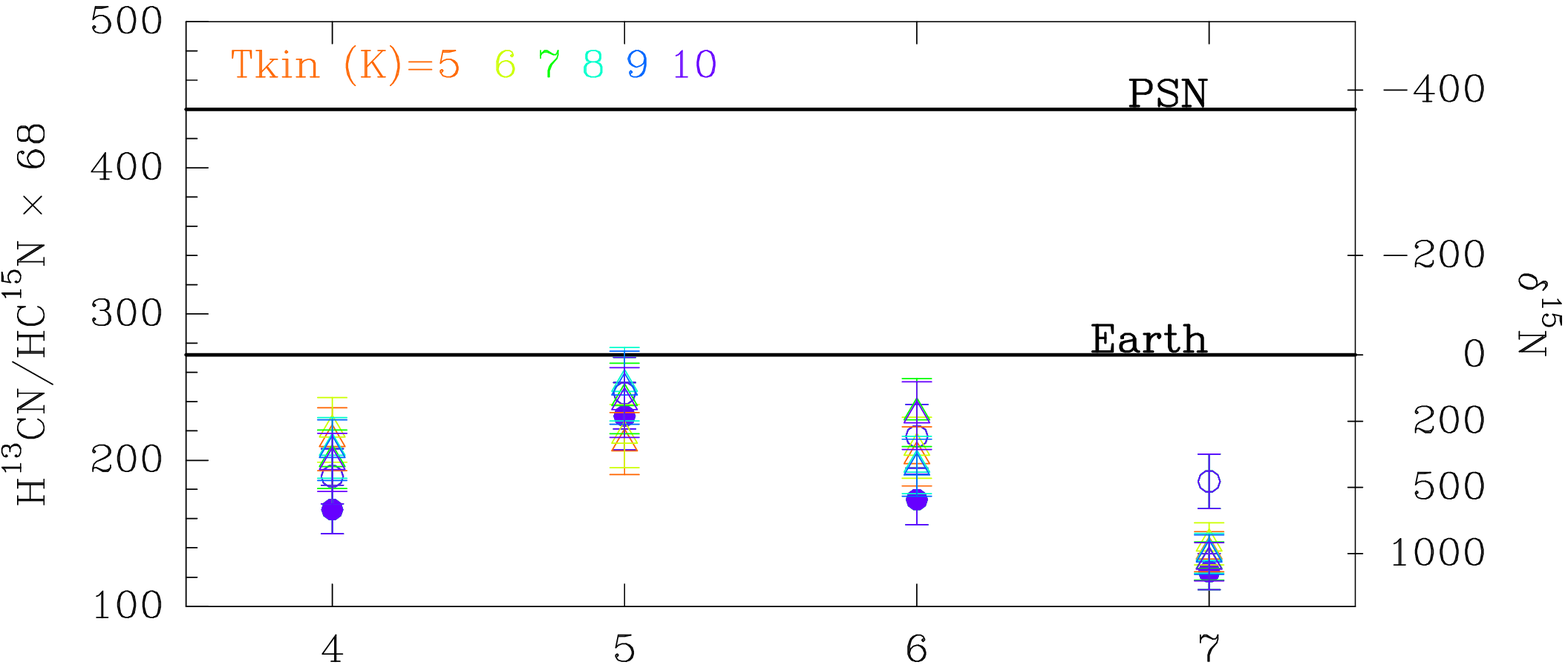}\\
  \includegraphics[width=\hsize]{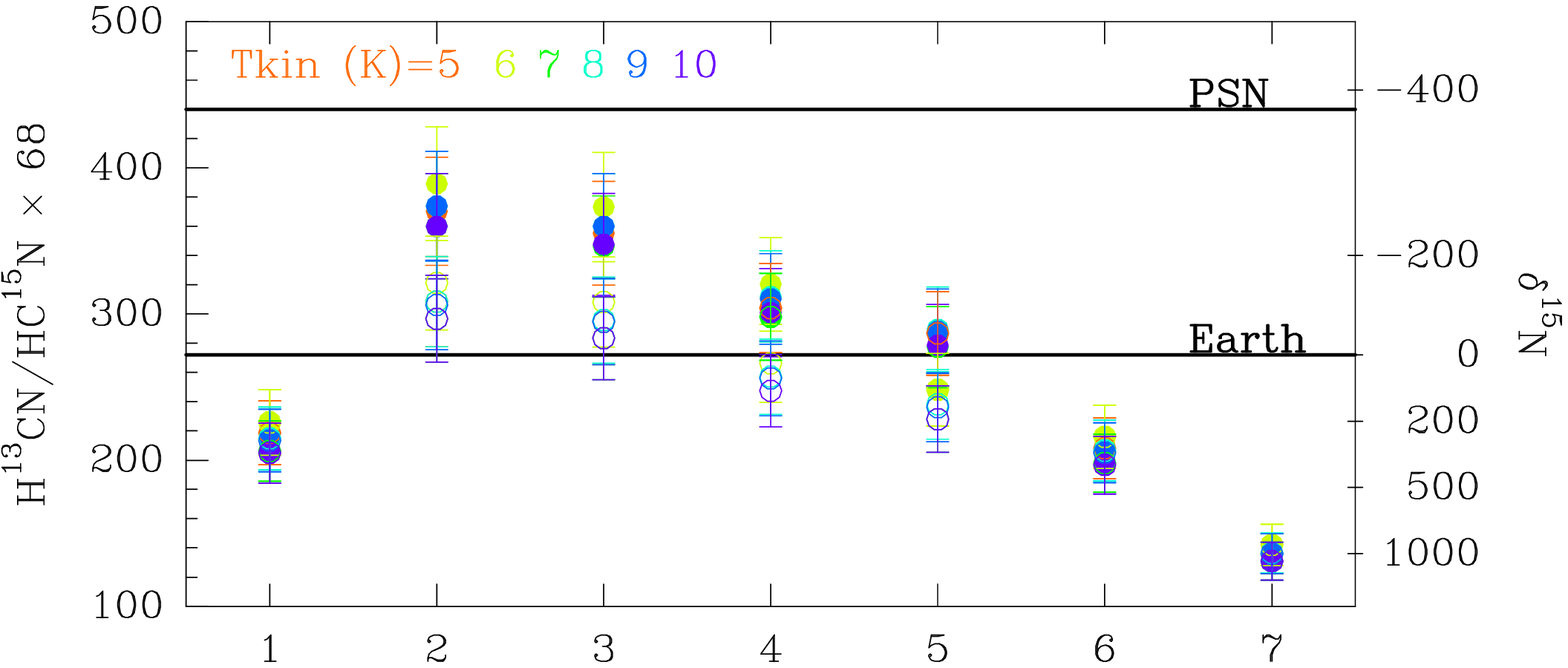}
  \caption{Nitrogen isotopic ratio \ratio{HCN} as measured towards
    L183 (upper panel) and L1544 (lower panel). Measurements for
    several positions are reported for each core (see also
    Tables~\ref{tab:l183} and ~\ref{tab:l1544}). The isotopic
    enrichment in delta notation is indicated on the right scale. In
    each panel, the thick lines indicate the protosolar nebula value
    of $441\pm 5$ \citep{marty2011} and the terrestrial reference
    ($\atoratio=272$). The ratios determined for kinetic temperatures
    ranging from 5 to 10~K are shown. At each position and for each
    kinetic temperature, several values are displayed, which
    correspond to different analysis methods (see Appendix for
    details).}
  \label{fig:observed-ratio}
\end{figure*}

The excitation temperatures are in the range 3--4~K, which is
significantly lower than the value assumed by \cite{hilyblant2010n},
but very close to the values determined by \cite{padovani2011} towards
other starless cores. The associated column densities lead to isotopic
ratios \hthcn/\hcfifn = 2 to 4.5.  As is evident from
Fig.~\ref{fig:observed-ratio}, the LVG column densities of both
\hthcn\ and \hcfifn\ depend only slightly on the kinetic
temperature. Within a given source, the range of values for the
isotopic ratio reflects genuine spatial variations accross the
source. These variations are up to a factor of 2 in L1544.

To derive the isotopic ratio \ratio{HCN} we assumed that
$[\ce{HCN}]/[\hthcn] = [\twc]/[\thc]$, such that
\begin{equation}
  \frac{[\ce{HCN}]}{[\hcfifn]} = \frac{[\hthcn]}{[\hcfifn]} \times
  \frac{[\twc]}{[\thc]}.
\end{equation}
\corr{This amounts in assuming that HCN does not undergo significant
  carbon fractionation and that the HCN/\hthcn\ ratio reflects the
  elemental ratio. Carbon fraction of HCN is unlikely for several
  reasons. First, most of the carbon is locked into CO and \thCO, and
  little carbon ions are then available for isotope exchange. In
  addition, \cite{milam2005} concluded that CN is at most only weakly
  affected by chemical fractionation, and the chemical similarity
  between CN and HCN led \cite{adande2012} to argue that carbon
  fractionation of HCN must be small. Last, it is to be noted that
  chemical fractionation would increase the \hthcn/HCN hence driving
  the molecular isotopic ratio \ratio{HCN} towards lower values. We
  thus argue that the nitrogen fractionation observed in HCN is
  robust.}  For the elemental isotopic ratio $^{12}$C/$^{13}$C of
carbon, we adopt the value of 68 from \cite{milam2005}. The values for
this ratio range from 140 to 360 towards L1544, and from 140 to 250
towards L183. These values are significantly lower than the isotopic
ratios reported by \cite{bizzocchi2010} towards L1544, using \ce{N2H+}
as a tracer. They are also well below the ratios determined towards
other cores by \cite{gerin2009nh3} and \cite{lis2010}, who used
\ce{NH2D} and \ce{NH3} as nitrogen carriers, respectively (see
Fig.~\ref{fig:ssratio}). In contrast, these values encompass the low
ratio $\ratio{HCN} = 150$ determined by \cite{ikeda2002} towards
L1521E.

In the following, we compare these results with isotopic ratios in
Solar System objects, and propose a unified view of these measurements
based on simple chemical arguments.

\section{Discussion}

\subsection{Differential fractionation for nitriles and amines}
\label{sec:network}

In prestellar cores, millimeter observations show that in contrast to
CO, nitrogen-bearing species such as CN and HCN manage to remain in
appreciable amounts in the gas phase \citep{hilyblant2008cn,
  padovani2011}. In such environments, isotope exchange reactions are
the only source of fractionation. These are caused by a thermodynamic
effect in which the exchange of isotopic atoms within a reaction has a
preferred direction owing to exothermicity, which is caused by zero
point energy differences. This process is efficient when the
temperature is lower than the exothermicity, provided that the
exchange reactions are competitive with other
reactions. \cite{rodgers2008} have shown theoretically that
significant \fifn\ enhancements can occur for various molecules in
N-rich prestellar cores depleted in CO and OH. As recognized by these
authors, however, their chemical model is hampered by the lack of
accurate rate coefficients for the numerous isotopologue exchange
reactions, which drive the fractionation.

Among the amines detected in prestellar cores, \ce{NH3} and its
deuterated isotopologues, and \ce{N2H+}, present isotopic ratios of
the order of 400 or larger \citep{gerin2009nh3, lis2010,
  bizzocchi2010}. In constrast, HCN shows significantly lower values
such as $\ratio{HCN}=150$ toward L1521E \citep{ikeda2002} and
$150-260$ towards L1544 \citep{milam2012}. These values are all
consistent with our new measurements towards L183 and L1544, also
based on HCN observations. Put all together, these observations
suggest a differential behaviour of nitriles and amines with respect
to fractionation (see Fig.~\ref{fig:ssratio} and
Table~\ref{tab:all}). This is indeed also visible in the
\corr{gas-phase} model of TH00, though at very low levels, \corr{and
  at a higher level in the gas-grain model of \cite{rodgers2008}.}

\begin{figure*}
  \centering
  \includegraphics[width=\hsize]{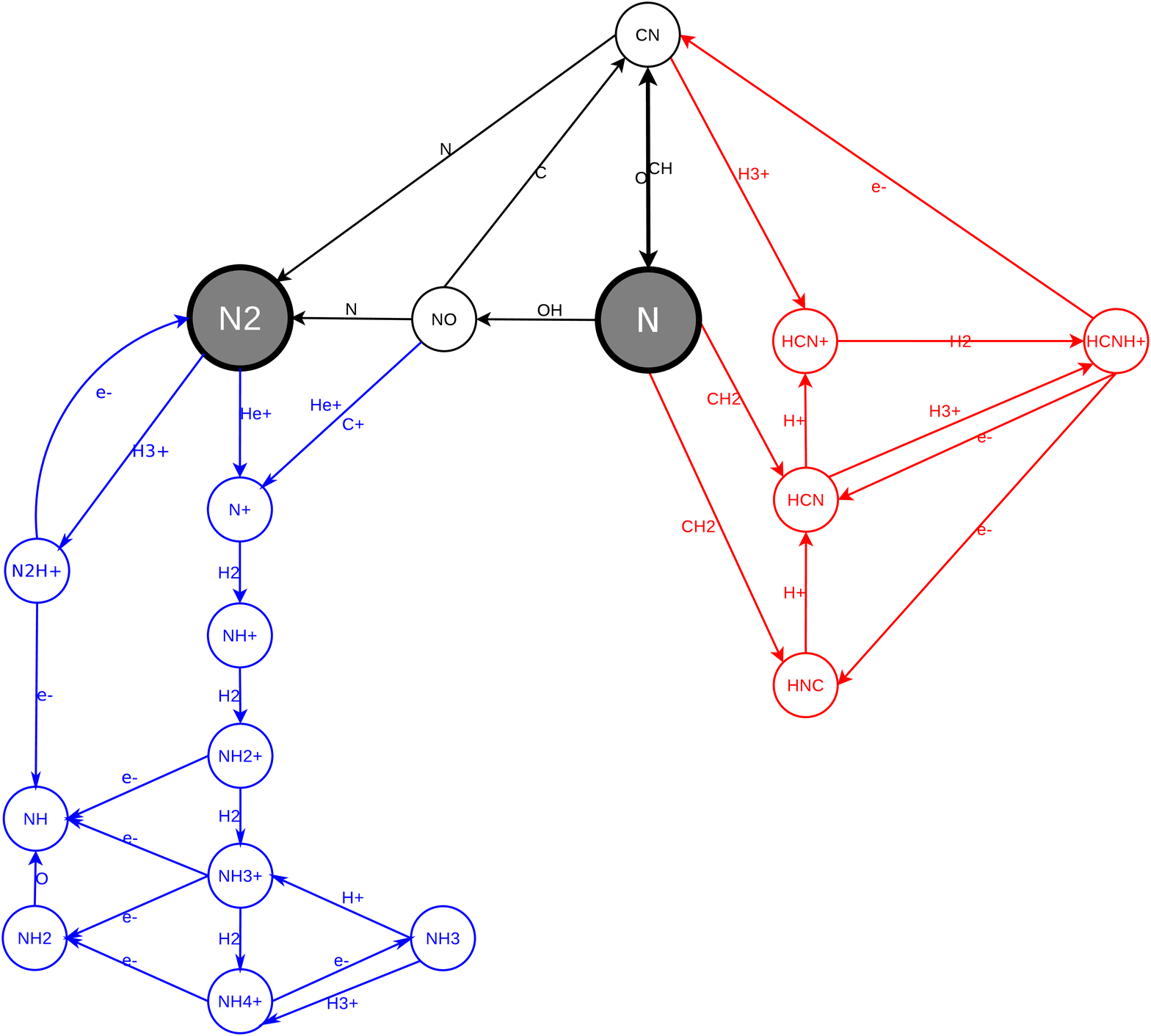}
  \caption{Principal gas-phase reactions involved in the interstellar
    chemistry of nitrogen in dense clouds where UV photons can be
    ignored . Amines (left) and nitriles (right) have been clearly
    separated.}
  \label{fig:network}
\end{figure*}

A comprehensive analysis of nitrogen interstellar chemistry in dark
clouds is summarized in Fig.~\ref{fig:network}. It appears that
N-bearing molecules can be divided into two almost distinct chemical
families: those carrying the nitrile (-CN) functional group and those
carrying the amine (-NH) functional group. The former family derives
from atomic nitrogen while the latter are formed \via\ \ce{N+}, which
is the product of \ce{N2} dissociative ionization. As a consequence,
these two families are not expected to exchange their \fifn. On the
other hand, the \atoratio\ exchange reactions among nitriles and
amines most likely present different time scales and/or efficiency
\citep[TH00, ][]{rodgers2008}.  Different \fifn\ enhancements are
therefore expected between \eg\ \ce{NH3} and HCN, even if rate
coefficients are uncertain.


The present work also shows that the nitrogen isotopic ratio varies
inside a given prestellar core. Although a variety of physical
parameters (density, temperature), known to present spatial variations
in these objects, could be invoked to explain these inhomogeneities,
source modelling including radiative transfer and chemistry is most
likely needed to draw conclusions in this regard. Yet, in the context
of this work, these spatial variations may be related with the large
range of values measured in the Solar System.

\begin{table*}
  \centering
  \caption{Nitrogen isotopic ratios in Solar System objects and in the cold ISM.}
  \begin{tabular}{l l @{\hskip5ex}c c l}
    \toprule
    Probe     & Source  & \ratio{}$\dag$  & \dearth$^\ddag$ &  References \\
    \midrule
                        
    \ce{NH2D} &Barnard 1& $470\pm150$ &$-420\pm180$   &  \cite{gerin2009nh3}\\
              &L1689B   & $810^{+600}_{-250}$ &$[-800:-500]$ & \cite{gerin2009nh3} \\
    \ce{NH3}  &Barnard 1& 334$\pm$50  &$-180\pm120  $ & \cite{lis2010} \\
    \ce{N2H+} & L1544   & $446\pm71$  &$-390\pm100$   & \cite{bizzocchi2010} \\
    \ce{HCN}  & L1521E  & 150       &      815     & \cite{ikeda2002} \\
              & L183    & [140: 250]&[1000:  80]   & This work \\
              & L1544   & [140: 360]&[1000:-245]   & This work \\
    \midrule
    Amino Acids        & & [263:230]& [37:184]    & \cite{sephton2002}\\
    IOM (bulk)         & & $< 195$   & 400          & \cite{alexander2007}\\
    IOM (hotspots)     & & $< 65$    & 3200         & \cite{busemann2006} \\
    Isheyevo - clasts  & & 50        & 4450         & \cite{bonal2010}\\
    IDPs (bulk)        & & [305:180]& [-107:  514] & \cite{floss2006}\\
    IDPs (hotspots)    & & up to 118 & 1300         & \cite{floss2006}\\   
    \bottomrule
  \end{tabular}
  \label{tab:all}
  \begin{list}{}
  \item$\dag$ Molecular isotopic ratio measured in a given N-bearing
    species.
  \item$\ddag$ Deviation from the standard terrestrial value in parts
    per thousand defined as $\dearth = [\ratio{\oplus}/\ratio{} -1
    ]\times 1000$, where $\ratio{\oplus} = 272$ is the nitrogen
    isotopic composition of the terrestrial atmosphere.
  \end{list}
\end{table*}

\subsection{Potential N reservoirs sampled by Solar System objects}

Similarly to interstellar clouds, the PSN was most likely composed of
several nitrogen reservoirs characterized by different relative
abundances and isotopic compositions. In interstellar clouds,
molecules carrying the nitrile functional group appear to be
systematically \fifn-enriched compared to molecules carrying the amine
group (see Section 1 and Fig.~\ref{fig:ssratio}). Thus, we propose
that the highly variable \atoratio\ ratios in objects of the Solar
System might simply reflect the interstellar nitrogen reservoir from
which they are originating. The Sun and giant planets, sampled atomic
and/or molecular nitrogen, considered as the major reservoir in the
PSN. Asteroids and comets, that are N-depleted compared to the Sun,
may have sampled minor, less volatile, and isotopically fractionated N
reservoirs of compounds such as HCN. These are found to be
systematically \fifn-enriched compared to the Sun, hence the PSN. In
the following paragraphs, we discuss in details each of these issues.

\subsection{Variable fractionation in Solar System objects}


Nitrogen isotopic composition was determined using ammonia in the
atmosphere of Jupiter \citep{fouchet2000, owen2001} leading to
$\ratio{NH3} \approx 440$. It is now largely interpreted as
representative of the average value for nitrogen in the solar
nebula. The similarity of the high Jovian and nebular \ratio{NH3} and
\ratio{N+} ratios, respectively, reinforces the ideas that molecules
carrying the amine functional group, deriving from \ce{N2}, are not
fractionated.


Comets may have better preserved than asteroids the volatile molecules
that were present in the protosolar cloud. The abundances of the
simple molecules such as CO, \ce{CO2}, \ce{CH3OH}, \ce{H2CO} and HCN
suggest indeed the partial preservation of an interstellar component
\citep{irvine2000}. These molecules, present as ices in the nucleus,
are detected in the coma after their sublimation when the comets
approach the sun. HCN is the most abundant N-bearing molecule that has
been detected so far (directly or through the CN radical), and also
the only one whose nitrogen isotopic composition was measured. There
has been some debate whether the radical CN is produced through the
photodissociation of HCN, or is a thermo-degradation product of
refractory CHON grains \citep{fray2005}. However, a genetic link
between HCN and CN is strengthened based in their comparable nitrogen
isotopic compositions deduced from a careful data reanalysis
\citep{bockelee2008} and on consistent production rates
\citep{paganini2010}. Therefore, a relevant comparison between
cometary molecules and ISM is possible through the same molecular
species: HCN. The CN nitrogen isotopic composition was measured in a
large number of Oort Cloud comets ($\rm ^{14}N/^{15}N_{\rm ave} = 144
\pm 6.5$) and Jupiter family comets, ($\rm ^{14}N/^{15}N_{\rm ave} =
156.8 \pm 12.2$) revealing a large \corr{and fairly constant} nitrogen
fractionation \corr{($130 < \atoratio < 170$)} \corr{with no
  dependence on the origin and heliocentric distance of the observed
  comets \citep[see the reviews by ][and references
  therein]{jehin2009, manfroid2009}}. These ratios are similar to the
lowest ones in dark clouds L1544 and L183 (see Fig.~\ref{fig:ssratio})
\corr{but do not reflect the spatial heterogeneity seen in these two
  dark clouds. However, these two clouds do not evidence the same
  levels of heterogeneity, and one possible explanation might be that
  the PSN emerged from a more homogeneous dark cloud than L183. In
  addition, only a small mass fraction of the material from the dark
  cloud -- that may be more homogenous -- is eventually incorporated
  stars and planetary systems. Last, potential isotopic heterogeneity
  of a cometary nucleus could remain undetected, since observations
  from the ground generally provide an averaged measurement of the
  coma \citep[\eg][]{blake1999}.} The N isotopic composition of HCN in
comets is therefore consistent with an interstellar
heritage. \corr{The preservation of cometary ices highly depends on
  the thermal history of the objects. The actual presence of the
  highly volatile HCN in comets attests of the absence of significant
  heating. Moreover, } only few processes are expected to modify the
nitrogen isotopic composition of the HCN molecule after its
formation. Indeed, nitrogen atoms are not easily exchangeable unlike
protons that easily exchange with ice. Evidences were provided
experimentally for protons in methanol \citep{ratajczak2009} and
through observations in HCN \citep{blake1999}. In this regard, the
\ratio{}\ ratio may appear as a more reliable proxy of the origin of
the molecule than the D/H ratio.


Chondrites might not be considered as a representative sampling of the
nitrogen in the PSN. Indeed, asteroids, hence chondrites, most likely
did not accrete the highly volatile nitrogen reservoirs (\ce{N2} and
N). Thus, the \fifn-enriched organics in chondrites might have
originally sampled some of the minor reservoirs made of nitrogen
compounds such as HCN and N-bearing molecules of higher molecular
weight.


Carbonaceous chondrites contain up to 5\% elemental carbon in a
variety of forms, organic matter being the major one. A minor fraction
(less than 25\%) of the organic matter in carbonaceous chondrites is
present as relatively low-molecular-weight compounds, extractable with
common organic solvents, the so-called ``soluble organic matter''
(SOM). SOM consists in a complex mix of organic molecules bearing H,
C, O, N, S, and P elements, with masses up to 800 amu
\citep{sephton2002, gilmour2003, schmittkoplin2010}. The remaining
fraction (75\% or so) is present as a high-molecular-weight
macromolecular material, persisting after harsh demineralization of
the chondrites, the so-called ``insoluble organic matter''
(IOM). Interplanetary Dust Particles (IDPs) and Antarctic
micrometeorites (AMMs) are micrometric particles that have either an
asteroidal or a cometary origin. They also contain organics that
present similarities with those of carbonaceous chondrites
\citep{dobrica2011}.

The soluble and insoluble organic fractions both contain some nitrogen
and are characterized by \fifn-enrichments relatively to the PSN. The
nitrogen isotopic compositions of amino acids have mostly been
determined in the Murchison chondrite \citep{pizzarello1994,
  engel1997}. The \ratio{} ratios are typically between 230 and
263. The isotopic fractionation is obviously reported on amine
functional groups that are not fractionated in our model scheme (see
Fig.~\ref{fig:network} and Sect.~\ref{sec:network}). The origin of
amino acids is yet unknown. Multiple pathways of formation have been
proposed in the literature. Some recent experiments on interstellar
ices analogs showed that a viable model of formation is based on
nitriles as amino acids precursor molecules \citep{elsila2007}. Hence
the nitrogen isotopic composition of amino acids might reflect that of
the precursor HCN and not that of \ce{NH3}.

Nitrogen is a minor element of IOM \citep[2\% in weight in average
][]{alexander2007}. It is mostly present in heterocycles such as
pyrroles \citep[\eg\ ][]{sephton2003, remusat2005, derenne2010}. The
contribution of N as present in nitriles appears to be relatively low
(\ce{N_{pyrrole}}/\ce{N_{nitrile}} = 5 in Murchison; Derenne and
Robert, 2010). The most primitive chondrites are characterized by bulk
\fifn-enrichments up to \ratio{} = 195
\citep{alexander2007}. Chondritic clasts in the unique Isheyevo
meteorite are characterized by the highest bulk \fifn-enrichment at
the present day, with \ratio{} = 50 \citep{bonal2010}. Analytical
techniques with submicron-scale imaging abilities revealed very
localized \fifn-enrichments (commonly referred to as \fifn-hotspots),
up to \ratio{} = 65 \citep{busemann2006}. Due to the experimental
challenges implied by their micron-scale size, SOM and IOM in IDPs are
not isolated; only isotopic compositions of bulk material are
measured. High \fifn-enrichments were revealed in IDPs; bulk such as
$180 < \ratio{} < 305$ - hotspots up to \ratio{} = 118
\citep{floss2006}. As a summary, similar \fifn-enrichments are
measured in bulk IOM of cosmomaterials and in HCN in L1544 and
L183. However, to be meaningful the comparison between ISM and
cosmomaterials must be based on similar molecules (\eg\ HCN in comets)
or on molecules linked by determined chemical pathways (\eg\ \fifn\ of
amino acids inherited from nitriles precursors). The chemical carriers
of the isotopic anomalies (bulk and hotspots) in the IOM are not
identified yet. They may be located onto heterocycles, nitriles,
and/or unidentified chemical group or compound. The IOM as currently
observed in cosmomaterials was most likely synthesized through
multistep processes that possibly involved recycling of interstellar
species within the protosolar disk \citep{sephton2002, dartois2004,
  okumura2011}. As a consequence, it is impossible to draw a direct
link between the \fifn-enrichments in the IOM of cosmomaterials to
interstellar molecules or to a series of chemical reactions as they
are expected to occur in ISM. Furthermore, physical processes like
radiolysis or heating could have modified IOM or even be involved in
its synthesis \citep[\eg\ ][]{huss2003}. Little is known about the
isotopic fractionation due to these processes, a significant role
cannot be excluded. Hence a genetic link between ISM molecules and IOM
cannot be currently firmly established, but is at least suggested
based on consistent \fifn-enrichments.

\section{Conclusions and perspectives}

\corr{Among the} arguments against the idea of \corr{interstellar}
chemistry at the origin of \fifn-enrichments in organics of primitive
cosmomaterials are: \textit{(i)} \corr{the assumption of} nitrogen
isotopic ratios of the order of 400 or higher in interstellar HCN and
\ce{NH3}; \textit{(ii)} \corr{the failure of classical gas-phase}
ion-molecule reactions in interstellar chemical models to produce
large \fifn-enrichments \corr{TH00}; \textit{(iii)} the absence of
spatial correlation between D- and \fifn-enrichments in primitive
organics is interpreted as a proof of different processes at their
origins \corr{\citep{briani2009, aleon2010, marty2010}}.

The observations reported here irrevocably show that considerable
nitrogen isotopic fractionation occurs at low temperature in the gas
phase of prestellar cores. These new measurements provide strong
constraints to interstellar chemistry models \corr{and are consistent
  with the early-time chemistry predicted by the gas-grain model of
  \cite{rodgers2008}}. Moreover, even though fractionation of both
hydrogen and nitrogen might reflect low-temperature gas-phase
chemistry, it is probably not driven by the same molecular
carriers. Indeed, the isotopic composition of a given species is
determined by the complex interplay of a reaction network and the
isotopic compositions of the precursors. In addition, the typical
exothermicities of reactions leading to D-enrichments and
\fifn-enrichments are different ($\approx 230~$K and $\approx$30K,
respectively) and leave room for a differential fractionation between
hydrogen and nitrogen, depending on the thermal history of prestellar
cores. \corr{Last, it was recently shown that varying the
  ortho-to-para ratio of \hh\ in interstellar chemistry can lead to
  D-enrichments and at the same time inhibit nitrogen fractionation
  \citep{wirstrom2012}.} There is thus little reason to expect
correlated isotopic anomalies between these two elements.

Even though the link between organics in primitive cosmomaterials and
interstellar molecules cannot be directly determined, isotopic
fractionation is a strong diagnostic feature. The present study
evidences that the large nitrogen fractionations observed in comets
and chondrites are consistent with a presolar chemistry. Several
arguments used against such an idea are here clearly invalidated.


\appendix

\section{Column density determination}
\label{sec:appendix}

For a resolved hyperfine structure spectrum, such as \hthcn(1-0), the
assumption of a common excitation temperature for all hyperfine
components allows a derivation of the excitation temperature and of
the opacity of each component. Radiative transfer through gas with a
constant excitation temperature leads to the following expression for
the emergent intensity in \texttt{ON-OFF} observing mode:
\begin{equation}
  \tmb = [J_\nu(\texc) - J_\nu(2.73)]\,(1-e^{-\tau}) = 
  \Delta J_\nu(\texc) \,(1-e^{-\tau})
  \label{eq:ratran}
\end{equation}
with $J_{\nu}(T) = T_0/[1-\exp(-T_0/T)]$ and $T_0=h\nu/k$. Noting
$r_k$ the relative intensities of the various components of a
hyperfine multiplet, the ratio of the opacities of two components is
$\tau_i / \tau_j = r_i / r_j$. Hence, assuming a constant \texc\ for
all hyperfine components of a given multiplet, one directly obtains
from Eq.~\ref{eq:ratran}, that
\begin{equation}
  \frac{\tmb_{, i}}{\tmb_{, j}} = 
  \frac{1-\exp(-r_i \tau_0)}{1-\exp(-r_j \tau_0)}
  \label{eq:hfs1}
\end{equation}
where we choose $\sum_i r_i = 1$ and we have noted $\tau_0 = \sum_i
\tau_i$. From the measured $\tmb$ and known $r_i$, it is thus possible to
derive $\tau_0$, from which the excitation temperature follows by
inverting Eq.~\ref{eq:ratran}. For lines of moderate opacity ($\tau_0 <
1$), peak or integrated intensity ratios may be used with no
difference. The column density is then obtained directly from the
integrated opacity of the hyperfine component $k$ as:
\begin{equation}
  \Ntot = \frac{8\pi\nu^3}{c^3}\,\frac{Q(\texc)}{A_kg_k}
  \frac{\int \tau_k(v) \ud v}{1-e^{-T_0/\texc}}  = {\cal N}_k(\texc)
  \int \tau_k(v) \ud v
  \label{eq:Ntot_hfs}
\end{equation}

\begin{figure*}
  \centering
  \includegraphics[width=\hsize]{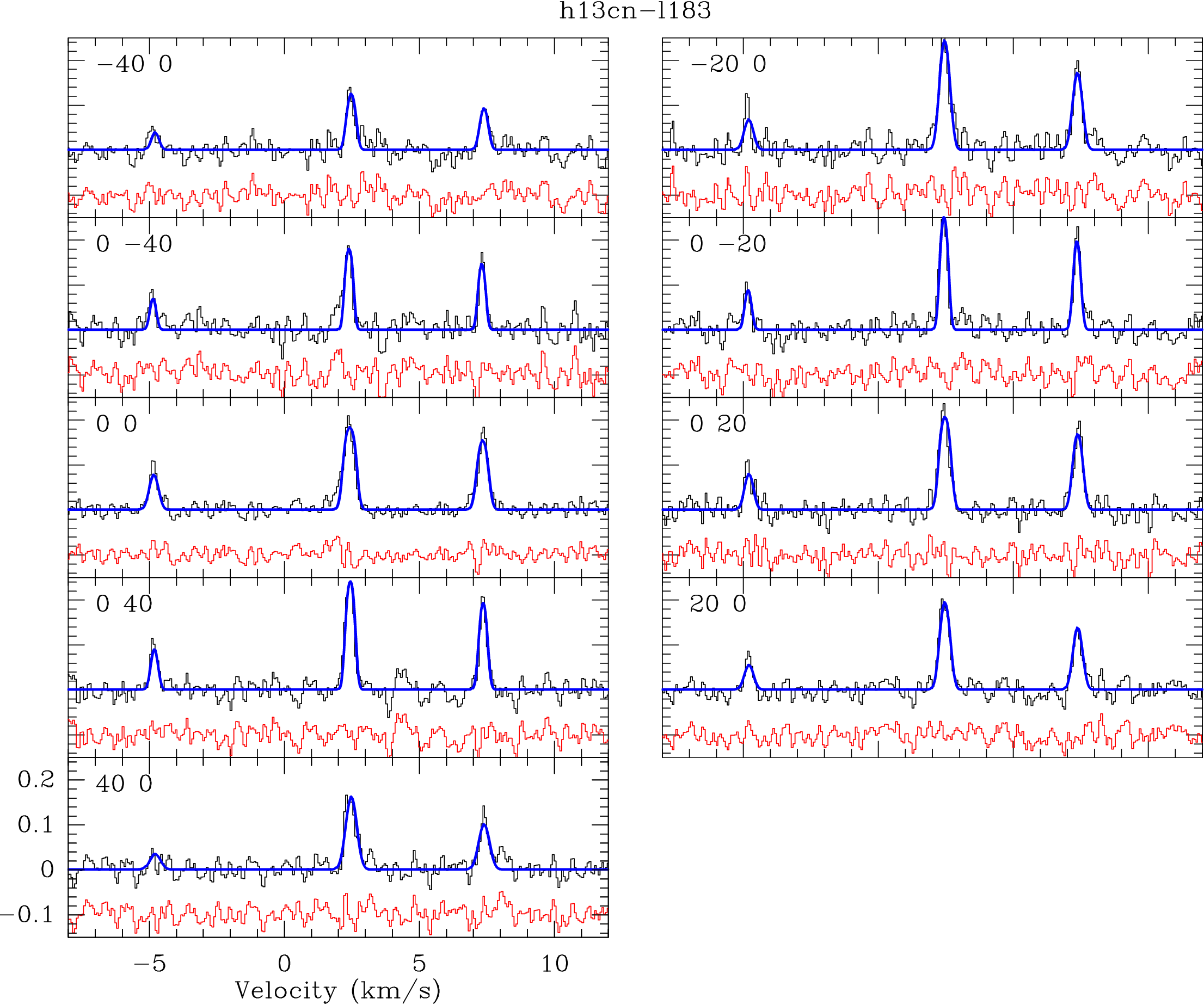}
  \includegraphics[width=\hsize]{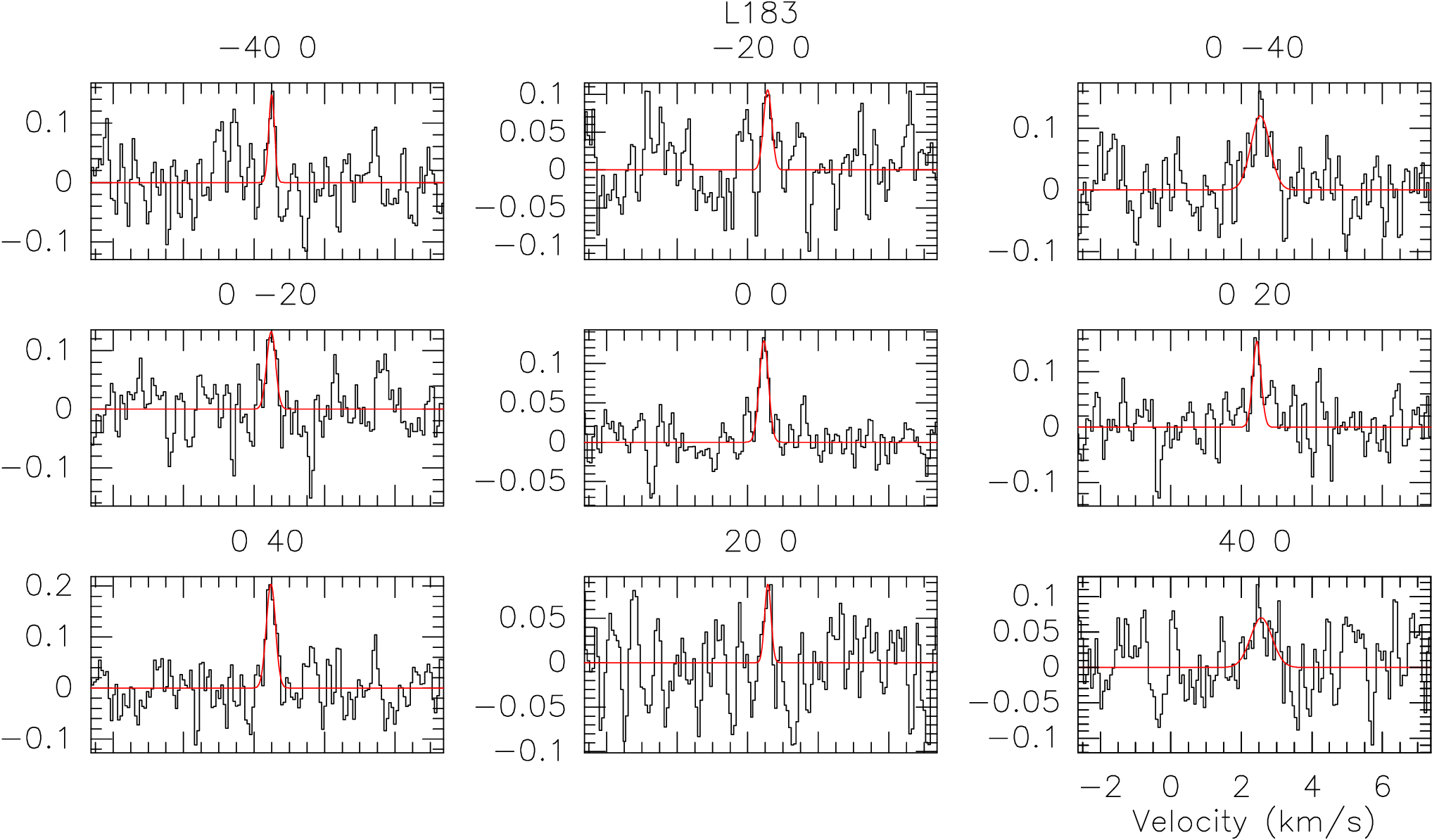}
  \caption{\textit{Top}: Results of the \ce{H^{13}CN}(1-0) hyperfine
    structure fitting procedure towards L183 for each spatial position
    (offsets in arcsec are indicated). The residuals are shown below
    the original spectrum. The fit result is overlaid on top of the
    spectrum. \textit{Bottom}: \hcfifn(1-0) spectra. Results from
    single-component Gaussian fits are shown.}
  \label{fig:spectral183}
\end{figure*}
\begin{figure}
  \centering
  \includegraphics[height=0.9\textheight]{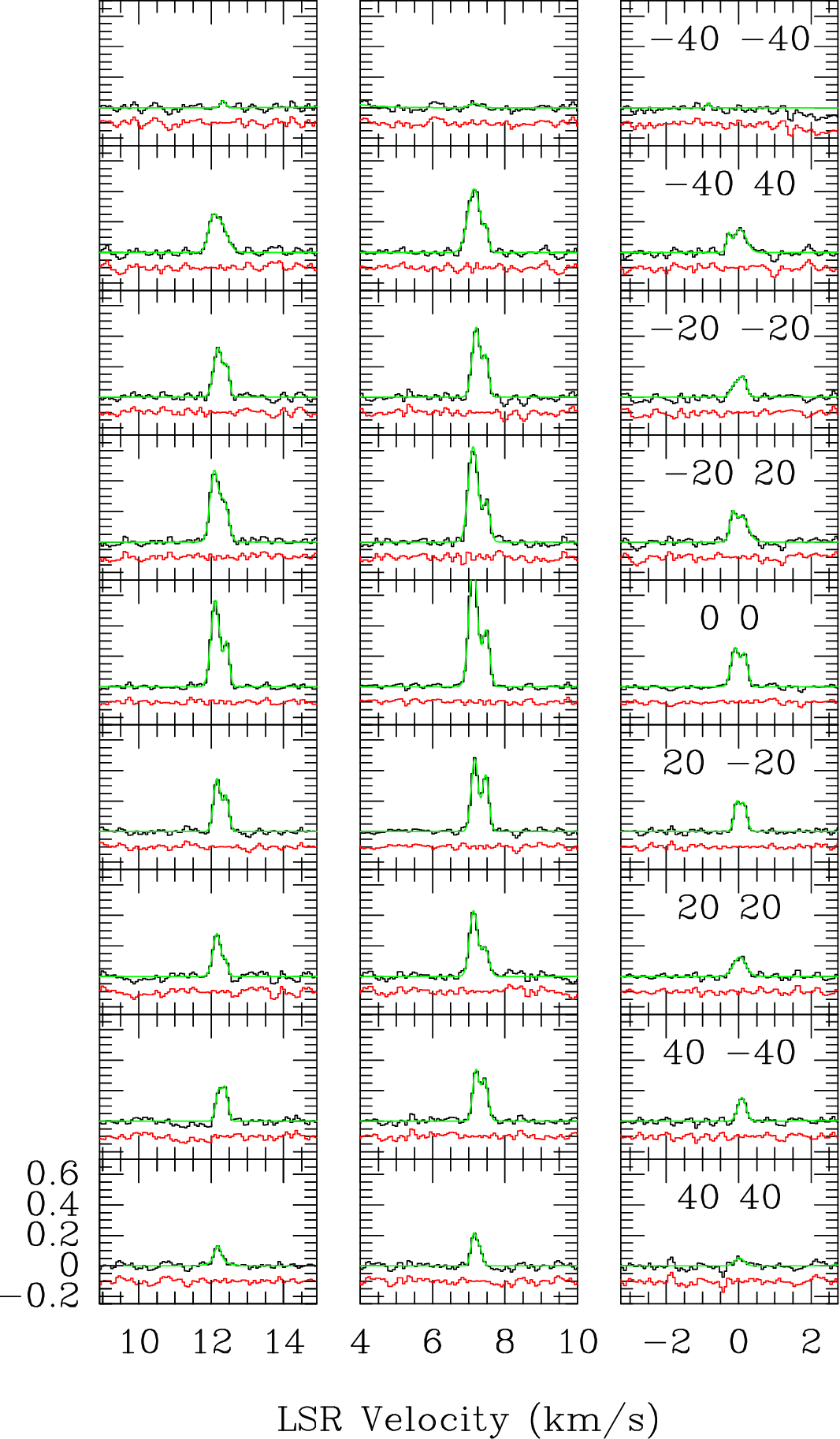}
  \caption{Results of the Gaussian fitting procedure applied to
    \hthcn(1-0) towards L1544 (see Sect.~\ref{sec:material}). On each
    row, the three hyperfine components are shown in separate panels
    emphasize the double peak profile. The residuals are shown below
    the original spectrum. The fit result is overlaid on top of the
    spectrum.}
  \label{fig:h13cn-l1544}
\end{figure}
\begin{figure*}
  \centering
  \includegraphics[width=\hsize]{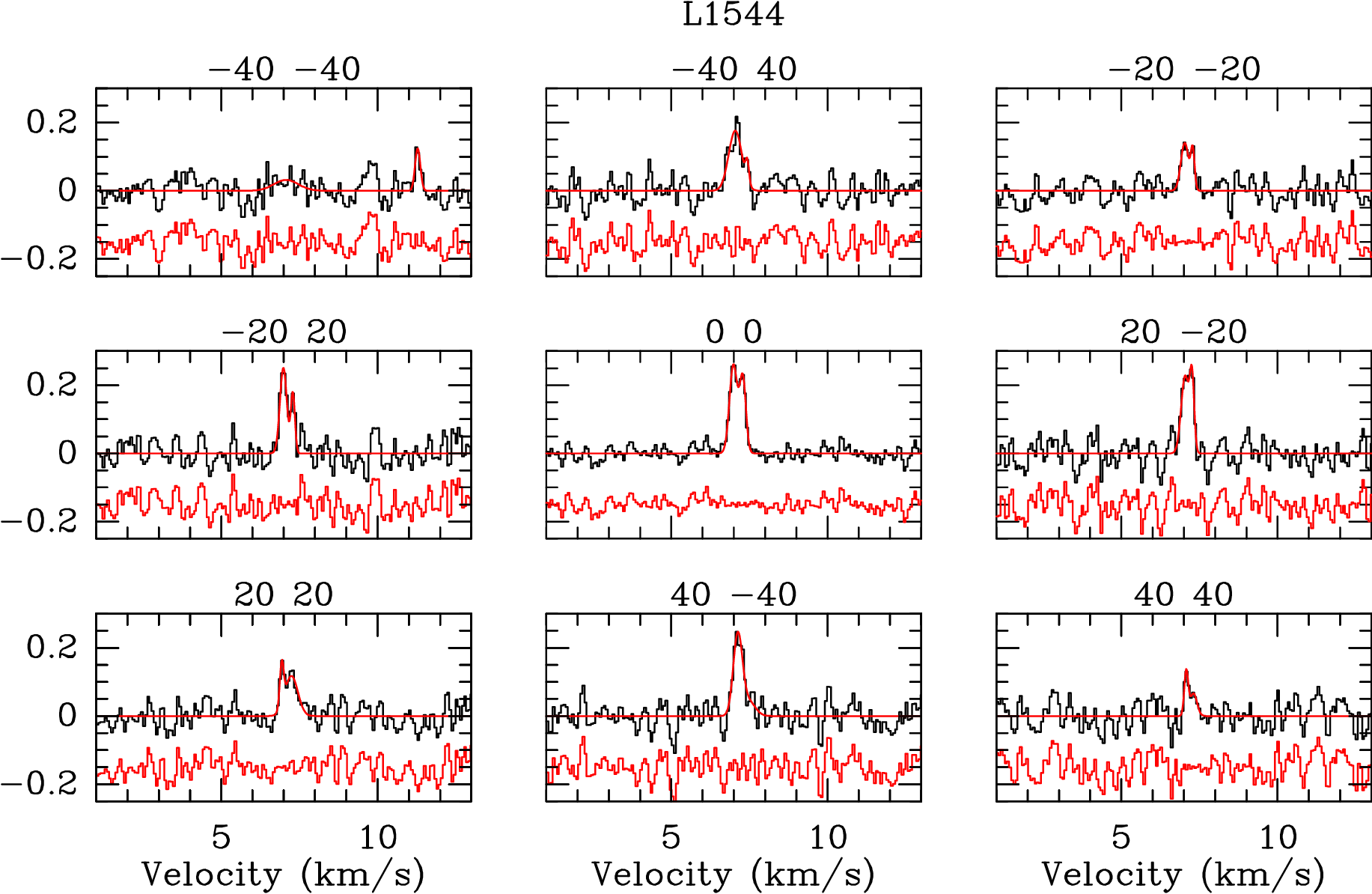}
  \caption{\hcfifn(1-0) spectra towards L1544. Gaussian fits are
    overlaid on top of each spectrum, and the residuals are shown
    below.}
  \label{fig:hc15n-l1544}
\end{figure*}

The HFS method of the CLASS software, used in the case of L183, fits
simultaneously the hyperfine components with Gaussians, by fixing
their relative positions and intensities. Fit results are shown in
Fig.~\ref{fig:spectral183}. In the case of L1544, the double-peak
nature of the emission spectrum made this procedure unfruitful. The
adopted strategy therefore was to first determine the integrated
intensity of each hyperfine component as obtained from independent
double-Gaussian fit (see Fig.~\ref{fig:h13cn-l1544} and
\ref{fig:hc15n-l1544}). The relative integrated intensities were then
used to derive the opacity and excitation temperature through
Eq.~\ref{eq:hfs1}.

\bibliographystyle{model2-names}
\bibliography{general,chemistry,phb,solarsystem}







\end{document}